\documentclass[sort&compress
    ,final            % use final for the camera ready runs
  ]
  {aipproc}

\layoutstyle{6x9}

\begin{document}

\title{Hadron formation in electron induced reactions at HERMES energies}

\author{T. Falter}{
  address={Institut fuer Theoretische Physik, Universitaet Giessen, 
D-35392 Giessen, Germany},
        ,email={Thomas.Falter@theo.physik.uni-giessen.de}
}
\author{W. Cassing}{
address={Institut fuer Theoretische Physik, Universitaet Giessen, 
D-35392 Giessen, Germany}
}
\author{K. Gallmeister}{
address={Institut fuer Theoretische Physik, Universitaet Giessen, 
D-35392 Giessen, Germany}
}
\author{U. Mosel}{
address={Institut fuer Theoretische Physik, Universitaet Giessen, 
D-35392 Giessen, Germany}
}

\begin{abstract}
We investigate meson electroproduction off complex nuclei in the kinematic regime of the HERMES experiment using a semi-classical transport model which is based on the Boltzmann-Uehling-Uhlenbeck (BUU) equation. We discuss coherence length and color transparency effects in exclusive $\rho^0$ production as well as hadron formation and attenuation of charged pions, kaons, protons and anti-protons in deep inelastic lepton scattering off nuclei.
\end{abstract}

\maketitle

%%%%%%%%%%%%%%%%%%%%%%%%%%%%%%%%%%%%%%%%%%%%
%% MAINMATTER
%%%%%%%%%%%%%%%%%%%%%%%%%%%%%%%%%%%%%%%%%%%%

High energy meson electroproduction off complex nuclei offers a promising 
tool to study the physics of hadron formation. The relatively clean 
nuclear environment of electron induced reactions makes it possible
to investigate the timescale of the hadronization process as well as the 
properties of hadrons immediately after their creation. In addition one can 
vary the energy and virtuality of the exchanged photon to examine the 
phenomenon of color transparency (CT). 

%%%%%%%%%%%%%%%%%%%%%%%%%%%%%%%%%%%%%%%%%%%%%%%%%%%%%%%%%%%%%%%%%%%%%%%%%%%%%%%%%%%%

In previous works \cite{Eff00,Fal02,Fal03a} we have developed a 
method to combine the quantum mechanical coherence in the entrance channel 
of photonuclear reactions with a full coupled channel treatment of the final 
state interactions (FSI) in the framework of a semi-classical transport 
model. This allows us to include a much broader class of FSI than usual Glauber theory. 

In our approach the lepton-nucleus interaction is split
into two parts: 1) In the first step the electron emits a virtual photon which is absorbed
on a nucleon of the target nucleus; this interaction produces a
bunch of particles that in step 2) are propagated within the
transport model. The virtual photon-nucleon
interaction itself is simulated by the Monte Carlo generator
PYTHIA v6.2 \cite{PYTHIA} which well reproduces the experimental
data on a hydrogen target. Instead of directly interacting with a quark inside the target nucleus the virtual photon might fluctuate into a vector meson ($\rho^0, \omega, \Phi, J/\Psi$) or perturbatively branch into a $q\overline{q}$ pair before the interaction. While the latter is very unlikely in the kinematic regime of the HERMES experiment as we have shown in Ref. \cite{Fal03a} the vector meson fluctuations become important at low $Q^2$ and clearly dominate the exclusive vector meson production measured at HERMES. The coherence length, i.e., the length that the photon travels as such a vector meson fluctuation $V$ can be estimated from the uncertainty principle:
\begin{equation}
\label{eq:coherence}
l_V=\frac{2\nu}{Q^2+m_V^2}.
\end{equation}
Here $\nu$ denotes the energy of the photon, $Q^2$ its virtuality and $m_V$ the mass of the vector meson fluctuation. If $l_V$ becomes larger than the internucleon distance in the nucleus the interactions triggered by the vector meson component $V$ get shadowed in nuclear reactions \cite{Fal02,Fal03a}.

A direct photon interaction or a non-diffractive interaction of one of the hadronic fluctuations leads to the excitation of one or more hadronic strings 
which finally fragment into hadrons. The time, that is needed for the
fragmentation of the strings and for the hadronization of the
fragments, we denote as formation time $\tau_f$ in line with the
convention in transport models. For simplicity we assume that the
formation time is a constant $\tau_f$ in the rest frame of each
hadron and that it does not depend on the particle species. 
We recall, that due to time dilatation the formation
time $t_f$ in the laboratory frame is then proportional to the
particle's energy
\begin{equation}
\label{eq:formation-time}
        t_f=\gamma\cdot\tau_f=\frac{z_h\nu}{m_h}\cdot\tau_f .
\end{equation}
Here $m_h$ denotes the hadron's mass and $z_h$ is the energy fraction of the photon carried by the hadron. The size of $\tau_f$ can be estimated by the time that the constituents 
of the hadrons need to travel a distance  of a typical hadronic radius
(0.5--0.8 fm). 

The formation time also plays an important role in the investigations of ultra-relativistic heavy ion reactions. For example, the 
observed quenching of high transverse momentum hadrons in $Au+Au$ reactions 
relative to $p+p$ collisions is often thought to be due to jet quenching in 
a quark gluon plasma. However, the attenuation of high $p_T$ hadrons might 
also be due to hadronic rescattering processes \cite{Gal02} if the
hadron formation time $\tau_f$ (in its rest frame) is sufficiently
short.

We assume that hadrons, whose constituent quarks
and antiquarks are created from the vacuum in the string
fragmentation, do not interact with the surrounding nuclear medium
within their formation time. For the leading hadrons, i.e. those
involving quarks (antiquarks) from the struck nucleon or the hadronic
components of the photon, we assume a reduced effective cross section 
$\sigma_{lead}$ during the formation time $\tau_f$ and the full hadronic 
cross section $\sigma_h$ ($h=\pi^\pm,K^\pm,p,\ldots$) later on.
The hadrons with $z_h$ close to one are predominantly leading hadrons and
can interact directly after the photon-nucleon interaction. Particles that 
emerge from the middle of the string might escape the nucleus due to time 
dilatation. However, about $2/3$ of these intermediate $z_h$ hadrons 
(mainly pions) are created from the decay of vector mesons that have been 
created in the string fragmentation. Because of their higher mass $m_h$ 
(0.77 -- 1.02 GeV) these vector mesons may form (or hadronize) inside the
nucleus (see Eq. (\ref{eq:formation-time})) and thus be subject to
FSI. The effect of the FSI, finally, will
depend dominantly on the nuclear geometry, i.e. the size of the
target nucleus.

The FSI are described by a coupled-channel
transport model based on the Boltzmann-Uehling-Uhlenbeck (BUU)
equation. For the details of the model we refer the reader to
Ref. \cite{Eff99}. The important 
difference to a purely absorptive treatment of the FSI is that the particles resulting from
the $\gamma^*A$ reaction do not have to be created in the primary $\gamma^*N$ interaction.
In a FSI with a nucleon a hadron might not only be absorbed but also be
decelerated in an elastic or inelastic collision. Furthermore, it
may in addition produce several low energy particles. In the case
of electroproduction of hadrons this finally leads to a redistribution
of strength from the high $z_h$ part of the hadron energy spectrum
to lower values of the energy fraction $z_h$. 

%%%%%%%%%%%%%%%%%%%%%%%%%%%%%%%%%%%%%%%%%%%%%%%%%%%%%%%%%%%%%%%%%%%%%%%%%%%%%%%%%%%%

\begin{figure}
  \includegraphics[height=.35\textheight]{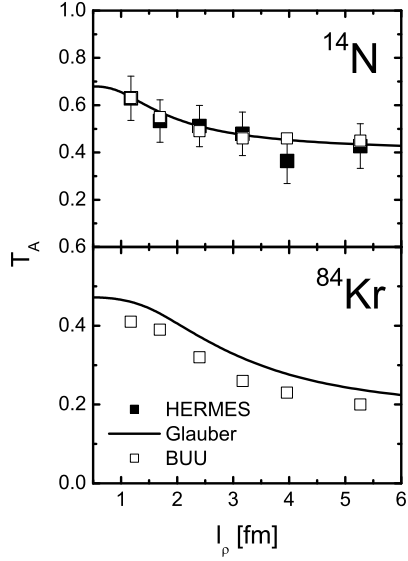}
  \caption{Nuclear transparency ratio $T_A$ (\ref{eq:transparency}) for $\rho^0$ electroproduction
plotted versus the coherence length (\ref{eq:coherence}) of the $\rho^0$ component of the photon.
The data is taken from~\protect\cite{HERMESrho}.
The solid line represents our Glauber result from Ref. \cite{Fal03a}. 
For each transparency ratio calculated within our 
transport model (open squares) we used the average value of $Q^2$ and $\nu$ 
of the corresponding data point.}
\label{fig:fig1}
\end{figure}

In Fig.~\ref{fig:fig1} we show the transparency ratio 
\begin{equation}
\label{eq:transparency}
  T_A=\frac{\sigma_{\gamma^*A\rightarrow \rho^0A^*}}{A\sigma_{\gamma^*N\rightarrow \rho^0N}}
\end{equation}
for exclusive $\rho^0$ production as a function of the coherence length 
(\ref{eq:coherence}) in comparison with the HERMES data \cite{HERMESrho}. The solid line displays the result that one gets if one 
uses our Glauber expression from Ref. \cite{Fal03a}. 

The result of the transport model is represented by the open squares. For 
each data point we have made a separate calculation with the 
corresponding values of $\nu$ and $Q^2$. For the $N$ target the Glauber and the 
transport calculation are in perfect agreement with each other and the 
experimental data. This demonstrates that, as we have discussed in 
Ref.~\cite{Fal02}, Glauber theory can be used for the FSI if the right 
kinematic constraints are applied. 

After applying all of the experimental cuts from Ref. \cite{HERMESrho}, nearly all of the detected $\rho^0$ 
stem from diffractive $\rho^0$ production for which we assume zero formation in the calculations. The $N$ data seems to support the assumption that the time needed 
to put the preformed $\rho^0$ fluctuation on its mass shell and let the wave 
function evolve to that of a physical $\rho^0$ is small for the considered 
values of $Q^2$. Furthermore, the photon energy is too low to yield a large 
enough $\gamma$ factor to make the formation length exceed the internucleon 
distance and make CT visible. This conclusion is at variance
with that reached in Ref.~\cite{Kop01} where the authors also stress that one might see an onset of CT when investigating the transparency ratio as a function of $Q^2$ for fixed coherence length.

We now turn to $Kr$ where we expect a stronger effect of the FSI. 
Unfortunately there is yet no data available to compare with. As can be seen
from Fig.~\ref{fig:fig1} the transport calculation for $Kr$ gives a 
slightly smaller transparency ratio than the Glauber calculation, especially 
at low values of the coherence length, i.e. small momenta of the produced 
$\rho^0$. There are two reasons for this: About 10\%~of the difference arises 
from the fact that within the transport model the $\rho^0$ is allowed to decay
into two pions. The probability that at least one of the pions interacts on 
its way out of the nucleus is about twice as large as that of the $\rho^0$. 
The other reason is that in the Glauber calculation only 
the inelastic part of the $\rho^0N$ cross section enters whereas the transport
calculation contains the elastic part as well. Thus all elastic scattering
events out of the experimentally imposed $t$-window are neglected in the
Glauber description. It is because of this $t$-window that also elastic 
$\rho^0N$ scattering reduces the transport transparency ratio shown in 
Fig.~\ref{fig:fig1}. Both effects are more enhanced at lower energies and 
become negligible for the much smaller $N$ nucleus.

%%%%%%%%%%%%%%%%%%%%%%%%%%%%%%%%%%%%%%%%%%%%
\begin{figure}
  \includegraphics[height=.5\textheight]{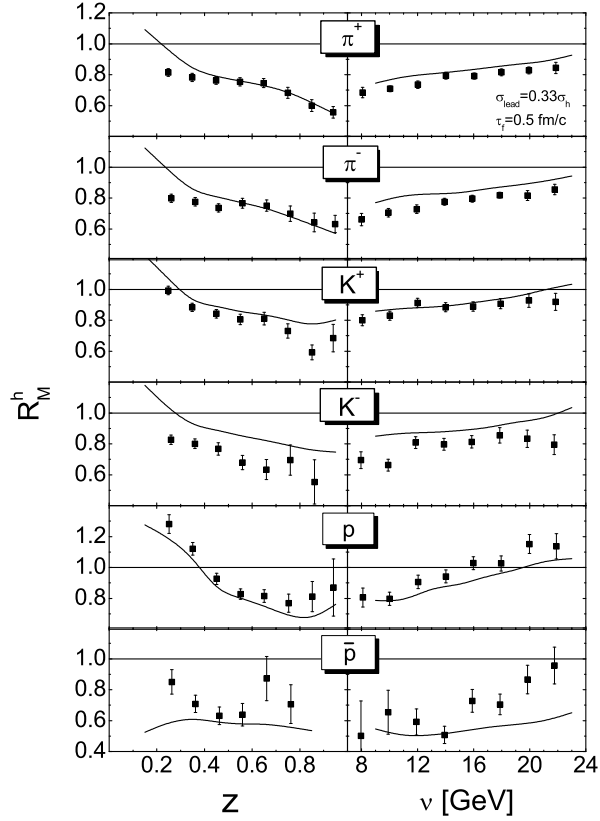}
  \caption{Calculated multiplicity ratios of $\pi^+$, $\pi^-$, $K^+$, $K^-$,
$p$ and $\bar{p}$ for $Kr$ using a fixed leading hadron cross section 
$\sigma_{lead}=0.33\sigma_h$ and formation time $\tau_f=0.5$~fm/c.
The experimental data has been taken from Ref. \cite{HERMES_new}.}
\label{fig:fig2}
\end{figure}

In Ref. \cite{Fal03b} we investigated the energy $\nu$ and fractional energy $z_h=E_h/\nu$
dependence of the charged hadron multiplicity ratio
\begin{equation}
\label{eq:multiplicity-ratio}
R_M^h(z,\nu)=\left(\frac{N_h(z,\nu)}{N_e(\nu)}\right)_A\bigg/\left(\frac{N_h(z,\nu)}{N_e(\nu)}\right)_D
\end{equation}
in DIS off nuclei and compared with the $N$ \cite{HERMESDIS_N} and $Kr$ \cite{HERMESDIS_Kr} data from the HERMES collaboration. In Eq. (\ref{eq:multiplicity-ratio}) $N_h(z,\nu)$
represents the number of semi-inclusive hadrons in a given ($z,\nu$)-bin and
$N_e(\nu)$ the number of inclusive DIS leptons in the same $\nu$-bin. 

In Ref. \cite{Acc02} the observed $R_M^h$ spectra were interpreted as being due to a combined effect of a 
rescaling of the quark fragmentation function in nuclei due to partial 
deconfinement as well as the absorption of the produced hadrons. Furthermore,
calculations based on a pQCD parton model \cite{Wang,Arl03} explain the 
attenuation observed in the multiplicity ratio solely by partonic multiple 
scattering and induced gluon radiation neglecting any hadronic FSI. It has already been pointed out by the authors of Ref. \cite{Acc02} that a shortcoming of the existing models 
is the purely absorptive treatment of the FSI. We avoid this problem by using the coupled-channel transport
model.

In Ref. \cite{Fal03b} we used the high $z_h$ part of the charged hadron data from Ref. \cite{HERMESDIS_Kr} to fix the leading hadron cross section to $\sigma_\mathrm{lead}=0.33\sigma_h$ during the formation time. The data for $N$ and $Kr$ could then be well described using a formation time $\tau_f >$ 0.3fm/c for all hadrons. This value is compatible with the analysis of antiproton 
attenuation in $p+A$ reactions at AGS energies \cite{Cas02}.

In Fig. \ref{fig:fig2} we show the results for the calculated multiplicity ratio of $\pi^-$, $\pi^+$, $K^-$, $K^+$, $p$ and $\bar{p}$ for {\it Kr} in comparison with the experimental data \cite{HERMES_new}. In our calculations we use the kinematic cuts of the HERMES experiment and take the detector geometry into account. We use a constant formation time of 0.5 fm/c and again scale all leading hadron cross sections with the same factor 0.33 during the formation time. Without further fine tuning we get a satisfying description of all the data meaning that the formation times of mesons, baryons and antibaryons are about equal.

The authors acknowledge valuable discussions with A. Accardi, N. Bianchi, A. Borissov, C. Greiner and V. Muccifora. This work was supported by DFG and BMBF.

%%%%%%%%%%%%%%%%%%%%%%%%%%%%%%%%%%%%%%%%%%%%%%%%%%%%%%%%%%%%%%%%%%%%%%%%%%%%%

%%%%%%%%%%%%%%%%%%%%%%%%%%%%%%%%%%%%%%%%%%%%%%%%%%%%%%%%%%%%%%%%%%%%%%%%%%%%

\end{document}